\documentclass[12pt]{article}  
\usepackage{amsmath}
\usepackage{url}
\usepackage{graphicx}
\usepackage{lscape}
\usepackage{rotating}
\usepackage{epstopdf}
\usepackage{setspace} 
\usepackage[font=footnotesize]{caption}
\usepackage{overcite}
\newenvironment{sciabstract}{%
\begin{quote} \bf}
{\end{quote}}
\usepackage[pagewise]{lineno} 

\newcounter{lastnote}

\title{Estimating the tolerance of species to the effects of global environmental change}
\author{Serguei Saavedra\footnote{To whom correspondence should be addressed. E-mail: serguei.saavedra@ebd.csic.es} \footnote{These authors contributed equally to this work} , Rudolf P. Rohr\footnotemark[\value{footnote}] , \\
Vasilis Dakos and Jordi Bascompte 
\\        
\\ Integrative Ecology Group \\ Estaci\'on Biol\'ogica de Do\~nana, EBD-CSIC \\ Calle Am\'erico Vespucio s/n, E-41092 Sevilla, Spain}

\date{}

\begin{document}
\maketitle
\baselineskip=8.5mm
 
\vspace{0.4 in}

{\bf Two-sentence summary:} Global environmental change is affecting the strength of interspecific interactions. The authors here estimate how much change species can tolerate before becoming extinct, and find that species tolerance is very sensitive to the net direction of change.

\newpage


\begin{sciabstract}
Global environmental change is affecting species distribution and their interactions with other species. In particular, the main drivers of environmental change strongly affect the strength of interspecific interactions with considerable consequences to biodiversity. However, extrapolating the effects observed on pair-wise interactions to entire ecological networks is challenging. Here we propose a framework to estimate the tolerance to changes in the strength of mutualistic interaction that species in mutualistic networks can sustain before becoming extinct. We identify the scenarios where generalist species can be the least tolerant. We show that the least tolerant species across different scenarios do not appear to have uniquely common characteristics. Species tolerance is extremely sensitive to the direction of change in the strength of mutualistic interaction, as well as to the observed mutualistic trade-offs between the number of partners and the strength of the interactions.
\end{sciabstract}

\clearpage

\section*{Introduction}

Global environmental change is accelerating as anthropogenic effects are increasing over short time scales\cite{Sala, Barnosky12}. The effects of such unprecedented change are modifying the abundance, physiology, and geographic range of individual species\cite{Thomas,Deutsch,Lorenzen}, and also affecting species interactions with the potential to modify ecosystem services, such as biological control and pollination\cite{Tylianakis,Burkle}. 

Empirical studies of mutualistic systems\cite{Tylianakis} suggest that the main drivers of global environmental change are nitrogen enrichment, increase in CO$_2$, and habitat fragmentation, among others. Importantly, these drivers can alter the frequency of pollinator visits to flowering plants due to climate-induced phenological shifts, nectar variability, or a decrease in flowers' abundance\cite{Tylianakis}. For example, it has been shown that solitary bees exhibit a small foraging range so that pollinator visitation can decrease if foraging distances become larger after habitat fragmentation\cite{Gathmann}. Similarly, it has been shown that land fragmentation can lead to a significant increase of pollinator visitation in a community of flower-visiting insects in red clover\cite{Diekotter}. In general, empirical evidence has shown that the effects of global environmental change can either increase or decrease the strength of mutualistic interaction, with the majority of cases showing the latter\cite{Tylianakis}.

While few studies have looked at the association of the effects of global environmental change with the loss of mutualistic interactions and community persistence\cite{Burkle}, species' tolerance to these effects have been restricted to pair-wise interactions\cite{Tylianakis}. Thus, it is still unknown the degree to which these effects will scale all the way up to entire networks of interactions, and which species would face a higher extinction risk.

On the theoretical front, recent work on ecological networks has shown important architectural properties that can facilitate species coexistence\cite{Bastolla,Thebault,Saavedra13}. Focusing on individual species, research has shown that species' generalization level\cite{Allesina} (i.e., its number of interactions, or degree) and their contribution to the nested architecture of the network\cite{Saavedra} play a key role for their survival. This work, however, has assumed constant environmental conditions.

Here, we introduce a theoretical framework to estimate species' tolerance to the effects of global environmental change. We focus on the association of species' tolerance with their level of generalization and their contribution to network architecture. To obtain a mechanistic understanding of the range of species' tolerance, we start by applying our framework to a 3-species community. We then  move to study species' tolerance in large communites. In general, we find that endangered species do not have unique characteristics. Importantly, our findings reveal that in order to estimate species' tolerance, first, one needs to identify the net effect of the observed environmental change.

\section*{Results}

\subsection*{A small mutualistic community}

We start by applying our framework (Methods) to a 3-species community (one plant and two pollinators with different strengths of mutualistic interaction represented by the width of links in Fig. 1a). Figures 1a and 1b show the community moving gradually from a weak to a strong mutualism, and from a strong to a weak mutualism, respectively. Note that in a weak mutualism regime, competition effects are stronger than mutualistic effects, whereas the opposite occurs in a strong mutualism regime. This implies that a species' initial abundance and intrinsic growth rate can be different depending on whether we start from a weak or a strong mutualism regime. We quantify a species' tolerance as the change in strength of mutualistic interaction that species can sustain before becoming extinct.

Under initial conditions, all species co-exist, but as soon as we start changing the strength of mutualistic interaction $\gamma_{o}$, species' initial abundances also begin to change (Fig. 1). While gradual changes in the strength of mutualistic interaction can increase the abundances of the plant and one of the pollinators, the interspecific competition between pollinators can drive one of the two into extinction. Interestingly, pollinators' tolerance is not the same under both directions of change. One pollinator (red) goes extinct when moving from weak to strong mutualism, but it survives when moving from strong to weak mutualism. The other pollinator (blue) presents the opposite pattern. This suggests that species' tolerance can be extremely sensitive to both the direction of change in the strength of mutualistic interaction and the variability of this strength across species.

To scale up from small to large communities, we apply our framework to 59 pollination and seed-dispersal networks (Methods). Figure 2 shows the species' tolerance (node color) of each of the 80 plants and 97 pollinators belonging to one network located in the high temperate Andes of central Chile\cite{Arroyo}. The darker the color, the higher the change in the strength of mutualistic interaction sustained by each species before becoming extinct and, in turn, the stronger its tolerance. In this example, changes are introduced by moving from a weak to a strong mutualism with a low mutualistic trade-off between a species' number and strength of interactions (Methods).

Figure 2 also shows the degree (node size) and contribution to nestedness (node position) for each of the species. Degree is given by the initial number of mutualistic interactions, while contribution to nestedness is measured by the extent to which those interactions contribute to the nested architecture of the network relative to an expected contribution (Methods). These two measures are very weakly correlated, providing almost independent information\cite{Saavedra}. Surprisingly, this figure reveals that some specialist species can be more tolerant than generalist species (far right). In the next sections, we explore how general this observation is and under what scenario is most likely to be found. We first consider the case where the effects of global change increase the strength of mutualistic interaction, followed by the reverse case where global change weakens the strength of mutualistic interaction, which seems to be the most likely case in nature\cite{Tylianakis}.

\subsection*{Increasing mutualism}

It has been shown that, under constant environmental conditions, generalists and weak contributors to nestedness can have the highest chances of survival\cite{Thebault,Saavedra,Allesina}. To test if this observation holds under an increase in the strength of mutualistic interaction (Methods), we use Spearman rank correlation to measure and properly compare across all networks the association of a species' tolerance with its degree and its contribution to nestedness. We find positive and significant correlations in both cases (Figs. 3a-3c). Interestingly, the positive association between species' tolerance and degree does not hold when the trade-off between the number of interactions and benefits received is large (Fig. 3c). In general, these results reveal that generalists and strong contributors to nestedness are the most tolerant of an increase in the strength of mutualistic interaction.

To further explore the magnitude of these associations, we calculate the ratio of the correlation norms $d$ of nestedness with species' tolerance and degree with species' tolerance across the 59 networks (Methods). The ratio measures how comparable the two correlations are in the magnitudes. A value of $d>1$ corresponds to the case where contribution to nestedness has on average larger correlations with species' tolerance than degree, and vice versa for values of $d<1$ (Fig. 3). We find a ratio of $d=1.16$ when the mutualistic trade-off is small (Fig. 3b), and ratios $d<1$ in the other two scenarios. This suggests that degree is not always the best estimator of species' tolerance. This also suggests that under a small level of mutualistic trade-off, the nested organization of the network can have the highest influence on the persistence of the community.

\subsection*{Decreasing mutualism}

More typically, however, global environmental change is expected to weaken the strength of interactions in mutualistic networks\cite{Tylianakis} (Methods). Surprisingly, we find the opposite patterns of what we found when moving from a weak to a strong mutualism. Here, generalists are highly tolerant with small and large mutualistic trade-offs (Fig. 3e-3f), while strong contributors to nestedness are the least tolerant in the three scenarios (Fig. 3d-3f).

Importantly, we find that with none and small mutualistic trade-offs (Fig. 3d-3e), species' tolerance is more strongly associated with their contribution to nestedness than with degree, i.e., $d>1$. These results, together with the ones found when increasing mutualism, show that the least tolerant species across different scenarios do not appear to have uniquely common characteristics. Instead, our results reveal that species' tolerance depends on both the direction of change and the mutualistic trade-offs.

\section*{Discussion}

In this paper, we have introduced a theoretical framework to estimate species' tolerance to change in the strength of mutualistic interaction as a potential effect of global environmental change. We have analyzed the extent to which both the direction of change and the mutualistic trade-offs are associated with species' tolerance. The former was investigated by moving from weak to strong mutualism and vice versa. The latter was investigated by modulating an observed mutualistic trade-off between a species' number and strength of interactions. We have found that specific combinations of direction of change and mutualistic trade-offs can have a different impact on species' tolerance. 

Contrary to the scenario of constant environmental conditions, where degree is the gold standard measure for estimating species' tolerance, here we have demonstrated that in a changing environment this is not always the case. In fact, consistent with empirical observations\cite{Burkle,Cameron}, generalist species can be the most vulnerable. We have found that generalists are the least tolerant under two scenarios: when the effect of global environmental change strengthens mutualism with a large mutualistic trade-off; and when the effect of global environmental change weakens mutualism with a very small mutualistic trade-off. This suggests that the tolerance of generalist species needs to be estimated relative to the direction of change, as well as to the mutualistic trade-off affecting specialist and generalist species. This is important since generalist species can have significant implications for the long-term functioning of ecosystems\cite{Pocock,Rezende}. Moreover, we have found that under half of the observed scenarios, species' tolerance can be more strongly associated with their contribution to nestedness than with degree. In general, this reveals that the least tolerant species across different scenarios do not appear to have uniquely common characteristics.

Because many complex systems are facing systemic risk, the results presented here are not only of relevance in ecology but beyond \cite{Saavedra09,Saavedra092,May09}. For example, in socio-economic systems, strong contributors to nestedness can be the most vulnerable to fail\cite{Saavedra}. Within our framework, this could be explained by a possible decoupling of cooperative interactions when moving from strong to weak mutualism and a lack of adaptation to these new conditions. Similar results can be observed in the banking sector, where financial institutions around the world are strengthening or weakening their cooperative interactions \cite{Scheffer,Haldane}. A valuable lesson from our results is that a node in these cooperative networks is never too big, too connected, or too peripherial to fail.

As new studies continue to confirm a dramatic loss of pollination systems around the world\cite{Aizen09,Kleinet,Burkle,Cameron}, it is becoming increasingly important to properly identify the characteristics of vulnerable species in these networks. Our findings reveal that in order to estimate species' tolerance to change, first, one needs to identify the net effect of environmental change, which depends on both species-level and network-level properties. Specifically, both the  direction of change in the strength of mutualistic interaction and the mutualistic trade-off highly modulate the ranking of species in terms of their tolerance to the effects of global environmental change.


\section*{Materials and Methods}

{\bf Data.} We investigate species' tolerance over 59 mutualistic networks that were located at different abiotic conditions around the world. This dataset is published in Rezende et al.\cite{Rezende}.
\\ \\
{\bf Framework.} In the text below, we describe in detail our proposed framework to study species' tolerance to the effects of global environmental change.
\\ \\
{\bf Species' tolerance.} We quantify a species' tolerance as the change in the strength of mutualistic interaction that it can sustain before becoming extinct. 
\\ \\
{\bf Model.} We model the dynamics of mutualistic systems composed of a set of plants and a set of animals (indicated by the upper indices (A) and (P)) using the same set of differential equations as in Bastolla et al.\cite{Bastolla}:

\begin{equation} \label{equ:ode}
\left\lbrace \begin{array}{c}
	\frac{dS_{i}^{(P)}}{dt} = S_{i}^{(P)}(\alpha_{i}^{(P)} - \sum_{j} \beta_{ij}^{(P)} S_{j}^{(P)} + \frac{ \sum_{j} \gamma_{ij}^{(P)} S_{j}^{(A)}} { 1 + h  \sum_{j} \gamma_{ij}^{(P)} S_{j}^{(A)} }) \\
	\frac{dS_{i}^{(A)}}{dt} = S_{i}^{(A)}(\alpha_{i}^{(A)} - \sum_{j} \beta_{ij}^{(A)} S_{j}^{(A)} + \frac{ \sum_{j} \gamma_{ij}^{(A)} S_{j}^{(P)}} { 1 + h  \sum_{j} \gamma_{ij}^{(A)} S_{j}^{(P)} })
\end{array} \right.
\end{equation}

The equations for pollinator populations can be written in a symmetric form interchanging the indices $(P)$ and $(A)$. Since there is no data to fully parametrize our dynamical system with meaningful biological information, we use a mean field approximation\cite{Bastolla} for the competition term (i.e., $\beta_{ii}=1$ and $\beta_{ij}=0.2$ if $i\neq j$) and we set the handling time $h = 0.1$. While these can be taken as arbitrary values, we find that our main conclusions are robust to the choice of different parameters.

The variables $S_{i}$ denote the abundance of species $i$. The parameter $\alpha_i$ represents the intrinsic growth rate, and $\gamma_{ij}$ denotes the strength of the mutualistic interaction between plants and animals. Simulations are performed by integrating the system of ordinary differential equations using the Matlab routline ode45. Species are considered extinct when their abundance density $S_i$ is lower than $100$ times the machine precision.

Consistent with empirical observations\cite{Turchin,Jordano}, initial abundances are set proportional to the number of interactions. We initialize all growth rates such that the initial abundances are at a feasible and stable equilibrium (see below). Finally, since we assume that species have no time to adapt\cite{Thomas}, all surviving species preserve their initial growth rates, interspecific competition and asymmetric benefits through the simulations.
\\ \\
{\bf Mutualistic trade-off.} Consistent with field observations\cite{Vazquez,Margalef}, we generalize the soft mean field approximation of Bastolla et al.\cite{Bastolla} by introducing explicitly a trade-off between mutualistic strength and species degree, which modulates the mutualistic trade-off between generalists and specialists: 

\begin{equation} \label{equ:gamma}
	\gamma_{ij} = \frac{\gamma_{o} y_{ij}}{k_{i}^{\delta}},
\end{equation}\\
where $\gamma_{o}$ represent the basal level of mutualistic strength, $k_{i}$ the degree of species $i$, and $y_{ij} = 1$ if species $i$ and $j$ interact and zero otherwise. The parameter $\delta$ modulates the trade-off. The higher the trade-off, the higher the strength (mutualistic benefit) perceived by specialists. In our simulations, we consider $\delta = 0$, $\delta = 0.5$, and $\delta = 2$ for none, small (sub-linear), and large (super linear) mutualistic trade-offs, respectively. Note that the case $\delta = 0$ is equivalent to the soft mean field approximation used in Bastolla et al.\cite{Bastolla}. 
\\ \\
{\bf Direction of change.} Changes in the strength of mutualistic interaction are modeled by either increasing or decreasing the mutualistic strength $\gamma_{o}$ in the dynamic model described above. The initial and final $\gamma_{o}$ for the increasing direction are set, respectively, to $\gamma_{o}=0$ and $\gamma_{o}=10\tau$. Similarly, the initial and final $\gamma_{o}$ for the decreasing direction are set, respectively, to $\gamma_{o}=3\tau$ and $\gamma_{o}=0$. Here, $\tau$ is the analytical threshold at which each network changes from a weak to a strong regime (see below). Changes to $\gamma_{o}$ are introduced in small steps when reaching a new equilibrium of abundances. We find no significant differences if changes in $\gamma_{o}$ are introduced at any point of our simulations. During any new simulation step, species' initial abundances are the final abundances of the previous step.
\\ \\
{\bf Weak and strong mutualism.} By definition, a mutualistic system is in a weak regime, if and only if the following $2\times 2$ block matrix 

\begin{equation} \label{equ:M}
	M = \left[ \begin{matrix}
		\beta^{(P)} & -\gamma^{(P)} \\
		-\gamma^{(A)} & \beta^{(A)}
	\end{matrix} \right]
\end{equation}\\
is positive definite (i.e., all eigenvalues of $M + M^{T}$ are positive). If that condition is not satisfied, then we say that the system is in a strong regime. This definition is a generalization to the non-symmetric case of the weak/strong concept introduced in Bastolla et al.\cite{Bastolla}. Intuitively, being in a weak regime means that mutualistic interactions are ``weaker" than competitive interactions. In the case of a fully-connected network without interspecific competition ($\beta_{ij} = 0$ for $i \neq j$) and the same strength of mutualistic interaction between all pairs of species, this condition is equivalent to the inequality derived in Bascompte et al.\cite{Bascompte06}. The condition of being in the weak regime is a stronger stability condition than the usual one based on the eigenvalues of the Jacobian matrix evaluated at an equilibrium point. Under the weak condition, any feasible equilibrium point (i.e., strictly positive abundance values that vanish the right side of the model equation) is automatically globally stable since it is possible to construct a Lyapunov function\cite{Goh}. Then, when the system enters the strong mutualism regime, with a handling time of $h = 0$, non-trivial fixed points are not any more granted to be stable. This means that for a large enough $\gamma_{ij}$ the system blows up. The only way to recover the stability of an equilibrium point in the strong regime is, first to have a positive handling time $h > 0$, and second that the abundances at the equilibrium point are large enough such that the system is locally stable (around that equilibrium point). In our framework, the transition from weak to strong mutualism is simply computed as a threshold of the basal level of mutualistic strength $\gamma_{o}$. Given a network, competition parameter values, and a trade-off value $\delta$, we can find a positive threshold, called $\tau >0$, such that if $\gamma_{o} < \tau$ the system is in the weak regime, and if $\gamma_{o} \geq \tau$ the system is in the strong regime. Note that this threshold is network, competition parameters, and trade-off dependent. For the same value of $\gamma_{o}$, a given network may be in the strong regime while another network can be in the weak regime.
\\ \\
{\bf Initialization of weak to strong mutualism.} For our simulations from weak to strong mutualism, we range the mutualistic strength $\gamma_{o}$ from $0$ to 10 times the threshold $\tau$, i.e., $\gamma_{o} \in [0, \dots ,10\tau]$. We initialize the system such that at $\gamma_{o} = 0$ it is at a globally stable equilibrium where all species have a positive abundance. Consistent with field observations\cite{Turchin,Jordano}, we choose this equilibrium point such that the abundances are proportional to the species degree $k_i$ and the total abundances of each species guild is equal to $5$, i.e., $S_{i}(\gamma_{o} = 0) = 5k_{i} / \sum_{i} k_{i}$. Note that the global stability of this potential feasible equilibrium is granted as we are in the weak regime. Finally, to ensure that this feasible point is also an equilibrium point, we have to choose the intrinsic growth rate such that

\begin{equation} \label{equ:alpha0}
 \alpha_{i} = \sum_{j} \beta_{ij} S_{j}(\gamma_{o} = 0).
\end{equation}
\\
{\bf Initialization of strong to weak mutualism.} For our simulations from strong to weak mutualism, we range the mutualistic strength $\gamma_{o}$ from $0$ to 3 times the threshold $\tau$, i.e., $\gamma_{o} \in [0, \dots , 3 \tau]$. We initialize the system such that at $\gamma_{o} = 3 \tau$ it is at a stable equilibrium where all species have positive abundances. As above, we also choose this equilibrium point such that the abundances are proportional to the species degree $k_i$, i.e., $S_{i}(\gamma_{o} = 3 \tau) = S_{o} k_{i} / \sum_{i} k_{i}$. However, since we are in the strong regime, this equilibrium point may not be stable for any value of the total abundance $S_{o}$. In particular, for a very low value of $S_{o}$, the equilibrium point may be unstable, and we only can recover its stability when $S_{o}$ has crossed a threshold. Then, the first task is to find this threshold in total abundance. For that, we first have to linearize the right side of the equations system. The linearized system around equilibrium abundances $\hat{S}_{i}$ is given by:

\begin{equation} \label{equ:lin}
\left\lbrace \begin{array}{c}
	\frac{dS_{i}^{(P)}}{dt} = S_{i}^{(P)}(\tilde{\alpha}_{i}^{(P)} - \sum_{j} \beta_{ij}^{(P)} S_{j}^{(P)} + \sum_{j} \tilde{\gamma}_{ij}^{(P)} S_{j}^{(A)}) \\
	\frac{dS_{i}^{(A)}}{dt} = S_{i}^{(A)}(\tilde{\alpha}_{i}^{(A)} - \sum_{j} \beta_{ij}^{(A)} S_{j}^{(A)} + \sum_{j} \tilde{\gamma}_{ij}^{(A)} S_{j}^{(P)}),
\end{array} \right.
\end{equation}\\
where $\tilde{\alpha}_{i}^{(A)} = \sum_{j} \beta_{ij}^{(A)} \hat{S}_{i}^{(A)} - \sum_{j} \gamma_{ij} \hat{S}_{j}^{(P)} / ( 1 + h \sum_{j} \gamma_{ij} \hat{S} _{j}) ^2$ and the linearized strength of mutualistic interaction $\tilde{\gamma}_{ij} = \gamma_{ij} / ( 1 + h \sum_{j} \gamma_{ij} \hat{S} _{j}) ^2$ (similar expressions hold for plants). Then, for the linearized system we can extract the linearized version of the $M$ matrix: 

\begin{equation} \label{equ:M}
	\tilde{M} = \left[ \begin{matrix}
		\beta^{(P)} & -\tilde{\gamma}^{(P)} \\
		-\tilde{\gamma}^{(A)} & \beta^{(A)}
	\end{matrix} \right]
\end{equation}

Note that the elements of $\tilde{M}$ are functions of the abundance values around which we have linearized the dynamical system. Our stability condition is that $\tilde{M}$, the linearized version of $M$, is positive definite, i.e., the system is locally stable in the weak regime. In our framework, we need to find which value of total abundance $S_{o}$, with $\hat{S}_{i} = S_{i}(\gamma_{o} = 3 \tau)$, makes the matrix $\tilde{M}$ be positive definite. We choose this exact value as starting point for the total abundance of the feasible point. Finally, to make that feasible point an equilibrium point, we have to choose the intrinsic growth rate such that

\begin{equation} \label{equ:alpha0}
 \alpha_{i} = \sum_{j} \beta_{ij} S_{j}(\gamma_{o} = 3 \tau) - \frac{ \sum_{j} \gamma_{ij} S_{j}(\gamma_{o} = 3 \tau)} { 1 + h  \sum_{j} \gamma_{ij} S_{j}(\gamma_{o} = 3 \tau) }
\end{equation}
\\
{\bf Contribution to nestedness.} Individual contribution to nestedness for each species or node quantifies the degree to which nestedness compares with the same value when randomizing just the interactions of that particular node\cite{Saavedra}. In calculating nestedness contributions, the interactions of a node are randomized according to the null model specified in Bascompte et al.\cite{Bascompte03}; we used 1000 random replicates. Here, nestedness is quantified using the measure proposed in Bastolla et al.\cite{Bastolla}, which is analytically linked to the dynamics of the mutualistic model. Other measures of nestedness and null models yield the same general results for the species-level analysis\cite{Saavedra}.
\\ \\
{\bf Ratio of the norms.} The ratio $d$ of the correlation norms between contribution to nestedness and degree, $x$ and $y$, is defined as $d=|x|/|y|$, where $|x|=\sqrt{\sum_i^m x_i^2}$, $|y|=\sqrt{\sum_i^m y_i^2}$. Here, $x_i$ and $y_i$ correspond to the Spearman rank correlations for each of the $m=59$ observed networks. The ratio $d$ provides a measure of the relative length of the correlations between contribution to nestedness and degree in an $m-$dimensional space. Typically, values within $0.9<d<1.1$ are considered significantly similar. The same general results are obtained if we use species' sum of the strength of mutualistic interaction instead of their number of interactions. Similarly, we find no significant association of the observed Spearman rank correlations with network connectance or size, confirming the comparability of our results across networks.
\\ \\
{\bf ACKNOWLEDGMENTS} We thank Alex Arenas and Jason Tylianakis for insightful discussions. Funding was provided by the European Research Council through an Advanced Grant (JB), CONACYT (SS), FP7-REGPOT-2010-1 program under project 264125 EcoGenes (RPR), and Rubicon grant NWO and a Marie Curie IEF-EU fellowship (VD).
\\ \\
{\bf Author contributions} All authors contributed extensively to the work presented in this paper.
\\ \\
{\bf Competing financial interests} The authors declare no competing financial interests.


\clearpage
\medskip

\renewcommand{\baselinestretch}{1.5}
{\small
\bibliographystyle{naturemag}
\bibliography{bibliography}

}

\clearpage


\begin{figure}[ht]
\centerline{\includegraphics*[width= 1.1 \linewidth]{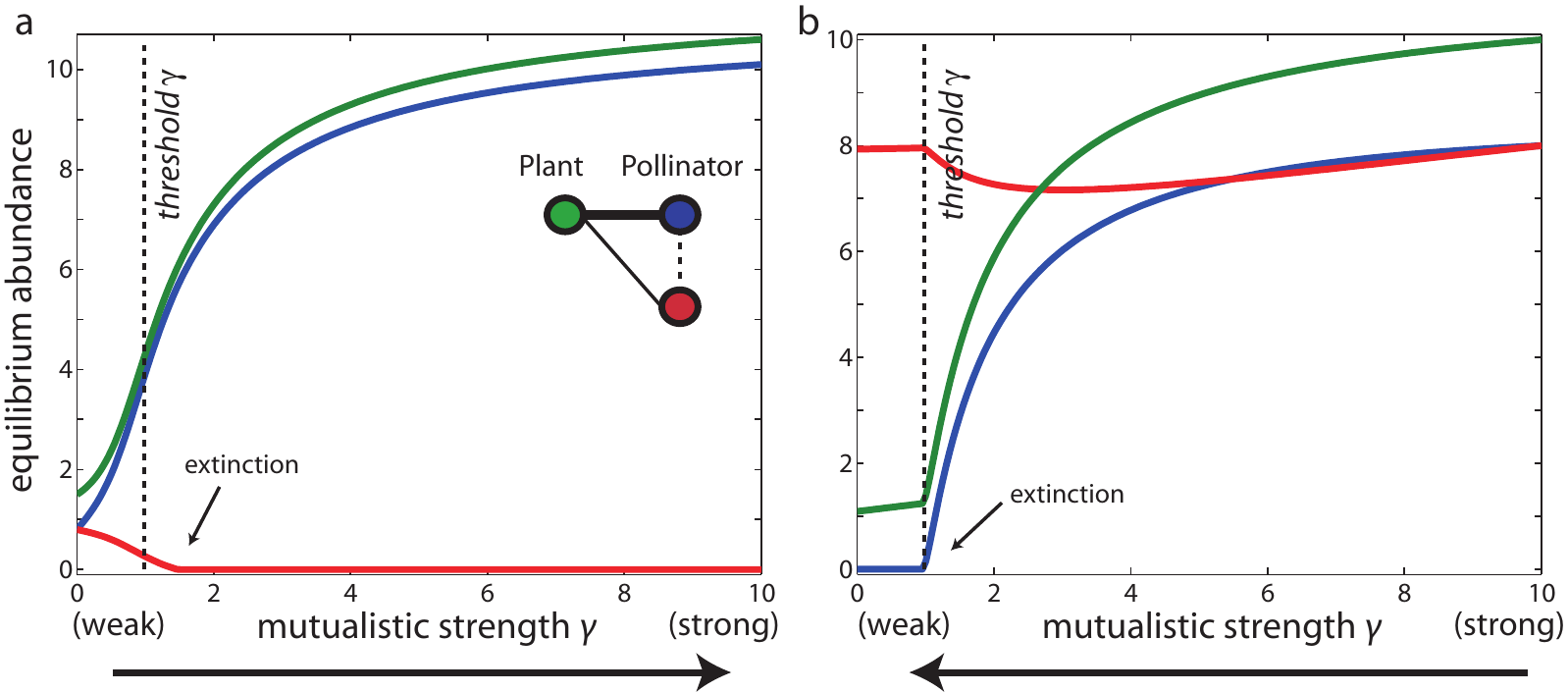}}
\scriptsize
\caption{\footnotesize Pollinators tolerance depends on the direction of change in the strength of mutualistic interaction. The figure plots the equilibrium abundances of a 3-species community (one plant and two pollinators) in the face of {\bf A} an increase and {\bf B} a decrease in the strength of mutualistic interaction. We set different strengths for each of the interspecific interactions (represented by the width of links). One pollinator (red) goes extinct when increasing the mutualistic interaction strength, but it survives when decreasing mutualism. The other pollinator (blue) presents the opposite pattern. The dashed line corresponds to the value of $\gamma_{o}$ at which the community shows a transition between weak and strong mutualism (Methods). In all our simulations we start at a feasible and stable equilibrium point (Methods).}
\label{fig1}
\end{figure}

\begin{figure}[ht]
\centerline{\includegraphics*[width= 1.1 \linewidth]{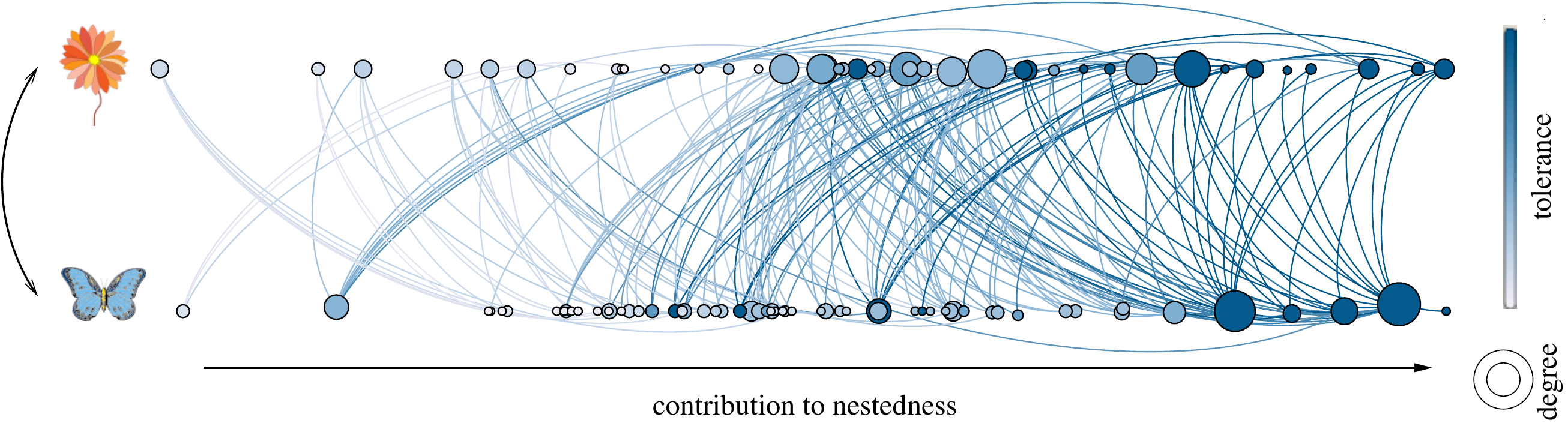}}
\caption{Species architectural characteristics and tolerance to change. The figure shows the species' tolerance (node color) of an increase in mutualism for each of the 80 plants and 97 pollinators belonging to one network located in the high temperate Andes of central Chile \cite{Arroyo}. The darker the color, the higher the change in the strength of mutualistic interaction sustained by each species before becoming extinct and, in turn, the stronger its tolerance. In this example, the effect of global change is assumed to increase the strength of mutualistic interaction. The system is simulated with a small mutualistic trade-off (Methods). We also show two key architectural properties of species: degree (node size) and contribution to nestedness (node position). Each symbol corresponds to one species or node and links correspond to the mutualistic interaction between plants (top) and pollinators (bottom). Interestingly, this figure reveals that some specialist species can be more tolerant than generalist species (far right).}
\label{fig2}
\end{figure}

\clearpage

\begin{figure}[ht]
\centerline{\includegraphics[width= 0.9 \linewidth]{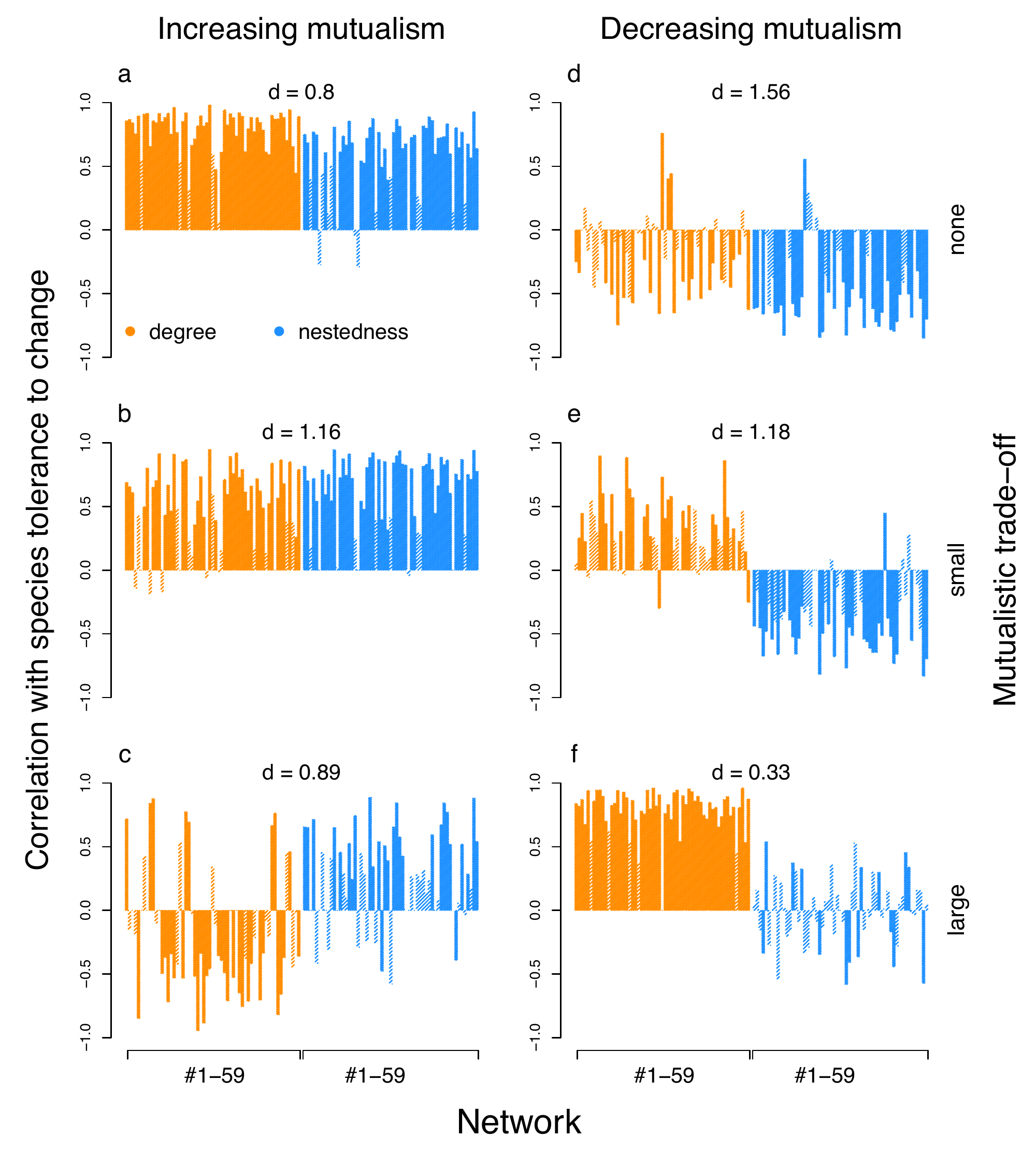}}
\caption{Species tolerance to change in large ecological networks. For each of the 59 networks (bars), the figure shows the Spearman rank correlation of animals' tolerance to change with degree (orange/left bars) and contribution to nestedness (blue/right bars). {\bf A-C} correspond to animals' tolerance to moving from a weak to a strong mutualism, while {\bf D-F} correspond to their tolerance to moving from a strong to a weak mutualism (Methods). Solid bars correspond to correlations that are significantly ($p<0.05$) different from zero. Correlations are calculated for different gradients of mutualistic trade-offs: {\bf A,D}, {\bf B,E}, and {\bf C,F} represent none, small, and large mutualistic trade-off, respectively (Methods). The figure also shows the ratio $d$ of the correlation norms between contribution to nestedness and degree (Methods). Plants' correlations are significantly similar to animals'. The figure reveals that in order to estimate species' tolerance, first, one needs to identify the correct direction of change and the mutualistic trade-offs in the system.}
\label{fig3}
\end{figure}

\end{document}